\documentclass[useAMS,usenatbib]{mn2e}
\usepackage{aas_macros,graphicx,times,multirow}

\newcommand{\vltn}{{\em Very Large Telescope}}
\newcommand{\vlt}{{\em VLT}}
\newcommand{\fors}{{\em FORS2}}
\newcommand{\forsn}{{\em FOcal Reducer/low dispersion Spectrograph}}
\newcommand{\xmm}{{\em XMM-Newton}}
\newcommand{\chan}{{\em Chandra}}
\newcommand{\ros}{{\em ROSAT}}
\newcommand{\ein}{{\em Einstein}}
\newcommand{\snr}{G315.4$-$2.3}

\title[Optical and X-ray observations of candidate isolated neutron stars in the \snr\ SNR]{Optical and X-ray  observations of candidate isolated neutron stars in the \snr\ SNR}
\author[R. P., Mignani, A. Tiengo, A. de Luca. ]{R.P. Mignani$^{1,2}$\thanks{E-mail:rm2@mssl.ucl.ac.uk}, A. Tiengo$^{3,4}$, A. de Luca$^{4,5}$\footnotemark[1]\thanks{Based on observations collected at ESO, Paranal, under Programme 385.D-0198(A)}\\
$^{1}$ Mullard Space Science Laboratory, University College London, Holmbury St. Mary, Dorking, Surrey, RH5 6NT, UK\\
$^{2}$ Kepler Institute of Astronomy, University of Zielona G\'ora, Lubuska 2, 65-265, Zielona G\'ora, Poland \\
$^{3}$ IUSS - Istituto Universitario di Studi Superiori, viale Lungo Ticino Sforza, 56, 27100, Pavia, Italy \\
$^{4}$ INAF - Istituto di Astrofisica Spaziale e Fisica Cosmica Milano, via E. Bassini 15, 20133, Milano, Italy\\
$^{5}$ INFN - Istituto Nazionale di Fisica Nucleare, sezione di Pavia, via A. Bassi 6, 27100, Pavia, Italy
}
\begin{document}

\date{Accepted 2012 June 28. Received 2012 May 18}

\pagerange{\pageref{firstpage}--\pageref{lastpage}} \pubyear{2012}

\maketitle

\label{firstpage}

\begin{abstract}
\snr\ is a young Galactic supernova remnant (SNR), 
whose identification 
as the remains of a Type-II supernova (SN) explosion has been debated for a long time. In particular, recent multi-wavelength observations suggest that it is  the result of a Type Ia SN, based on spectroscopy of the SNR shell and the lack of a compact stellar remnant.
However, two X-ray sources, one detected by \ein\ and \ros\ (Source V) and the other by \chan\ (Source N)
have been proposed as possible isolated neutron star candidates. In both cases, no clear optical identification was available and, therefore, we performed an optical and X-ray study  to determine the nature of these two sources. Based on
\chan\ astrometry, Source V 
is associated with a bright $V\sim14$ star, which had been suggested based on the 
less accurate \ros\ position. Similarly, from \vltn\ (\vlt) archival observations, we found that Source N is associated with a relatively bright star ($V=20.14 $).
These likely identifications suggest that both X-ray sources cannot be isolated neutron stars.
\end{abstract}


\section{Introduction}

The Galactic supernova remnant (SNR) \snr\  is a very young object, supposed to be the remains
of the historical  supernova SN 185 A.D. (Clark \& Stephenson 1975).  The remnant 
distance  is uncertain, with values ranging from 1 to 2.8 kpc (Kaastra et al.\ 1992; Rosado et al.\ 1996; Sollerman et al.\ 2003).   \snr\  
has been observed in radio (Hill 1967), in the soft X-rays by \ros\  (e.g., Vink et al.\ 2000), and hard X-rays by {\em ASCA} (e.g,  Borkowski et al.\ 2001) and,  more recently, both by \xmm\ and \chan\  (Vink et al.\ 2006).  In the optical, the southwest edge of \snr\ is associated with the RCW\, 86 nebula (e.g., Smith 1997).
At higher energies, 
it has been detected 
at TeV $\gamma$-rays by {\em H.E.S.S.} (Aharonian et al.\ 2009).
Its morphology is typical of a young shell-type SNR, but with a marked asymmetry in the X-ray surface brightness profile between the fainter northeast and the brighter southwest edges (Vink et al.\ 2000), as observed both in radio (Dickel et al.\ 2001) and in the optical, where only the region associated with the RCW\, 86 nebula is visible (Smith et al.\ 1997).  
This might imply an off-centred cavity SN explosion, with its centre closer to the southwest edge of the SNR and near the centre of the RCW\, 86 nebula. 
Recently, Williams et al.\ (2011) suggested that \snr\ is a Type-Ia SNR, based on optical and X-ray spectroscopy of the remnant and the lack of a compact source at the centre, which could be identified as the isolated neutron star (INS) formed out of a Type II SN explosion. 
In the radio band, Kaspi et al.\ (1996) could not identify a radio pulsar within the SN shell. In the X-rays, 
Vink et al.\ (2000) spotted an unresolved source (hereafter Source V) $\sim 7\arcmin$ southwest of the SNR centre, in the direction of the RCW\, 86 nebula.
However, the possible association with a $V\sim 14$ star coincident with the \ros\ 5\arcsec\ error circle and evidence for long-term variability suggested that this X-ray source was not an INS, although only a better localisation from X-ray observation at higher spatial resolution can either confirm or refute the association. More recently,  Gvaramadze \& Vikhlinin (2003)  
found two X-ray sources with \chan, located on the X-ray bright southwest edge of the \snr\ SNR.
Only the northernmost 
one (dubbed Source N)  had no optical counterpart ($R_F\ga 21$) and was proposed as the  stellar remnant of the SN explosion. However, 
its  possible identification 
as an INS has never been confirmed so far. 
Source N has not been detected in radio. 
One possibility would be that it 
is a radio-silent INS, possibly 
a Compact Central Object, or CCO (De Luca\ 2008).  A search for a stellar remnant in the central part of the SNR was carried out with \chan\ by Kaplan et al.\ (2004), who identified all the detected X-ray sources  either as foreground stars or background AGNs.  Thus, 
Source V and N remain the only unidentified X-ray sources potentially associated with the remnant of the SN explosion.
We investigated their nature 
using archival \chan\ data to better determine their positions, verify the association of Source V with the bright nearby star, and search for Source N's optical counterpart in archival \vltn\ (\vlt) data.

\section{Optical observations}

Optical images of the RCW\, 86 nebula,  at the southwest edge of  \snr\ (Fig.1, top left),  were obtained in service mode on April, 10 to 12 2010 with the \vlt\ Antu telescope at the ESO Paranal observatory\footnote{\texttt{www.archive.eso.org}}. 
The observations were performed with the \forsn\  (\fors) camera (Appenzeller  et  al.\ 1998), equipped with its  MIT detector, a mosaic of two 2k$\times$4k CCDs optimised  for wavelengths  longer  than 6000  \AA.   The standard low gain, fast read-out mode,  and the high-throughput  $v_{\rm HIGH}$ ($\lambda=5570$ \AA;  $\Delta \lambda=1235$\AA) filter were used. Images were taken with the standard resolution collimator (0\farcs25/pixel) and a field--of--view  (FOV) of 8$\farcm3  \times 8\farcm3$.
However, due to vignetting the effective sky coverage is smaller, and larger for the upper CCD chip  ($\sim 7\arcmin \times 4\arcmin$) than for the lower one.   Source N 
was positioned in the upper CCD chip, 
$\sim 30\arcsec$ above the gap between the two chips.  A total of 45 exposures of 300 s each were taken
under  dark time  and photometric conditions
with an average airmass of $\sim 1.3$ and image quality of $\sim 0\farcs6$.
We reduced the science images using standard tools in the {\sc iraf} package {\tt ccdred} for  bias subtraction and  flat--field correction. We then aligned and average-stacked the science images using the {\sc swarp} program (Bertin et al.\ 2002), applying a $3  \sigma$ filter  to reject hot/cold pixels and cosmic ray hits.  Since all exposures were taken with comparable image quality, we did not apply any selection prior to the image stacking.  We applied the  photometric calibration by using  the extinction-corrected night  zero points  computed by  the  \fors\ pipeline
derived from observations of the Landolt's standard stars (Landolt 1992).
We re-computed the  astrometric solution of the \fors\ image using stars selected from the Guide Star Catalogue 2 (GSC-2; Lasker et al.\ 2008).
We selected stars evenly distributed in the FOV  but far from the CCD edges and the vignetted regions, where geometric distortions are larger. We  measured the star centroids through Gaussian fitting  using the Graphical  Astronomy  and   Image  Analysis  ({\sc gaia})  tool\footnote{\texttt{star-www.dur.ac.uk/$\sim$pdraper/gaia/gaia.html}}  and used  the code  {\sc  astrom}\footnote{\texttt{www.starlink.rl.ac.uk/star/docs/sun5.htx/sun5.html}} to compute  the pixel-to-sky coordinate  transformation.
The  rms  of the astrometric  fit was $\sigma_{\rm r} \sim 0\farcs21$, in  the radial direction.   To this  value  we added  in  quadrature the  uncertainty $\sigma_{\rm tr}=0\farcs1$ on the registration of the \fors\ image on the  GSC2  grid, $\sigma_{\rm tr}= \sqrt{3/N_{s}} \sigma_{\rm  GSC2}$ (Lattanzi et al.\ 1997),  where  $\sigma_{\rm  GSC2}=0\farcs3$ is  the  mean GSC2 positional error and $N_{\rm s}=22$ is  the number of stars used  to compute the astrometric solution.  After accounting for the $\sim 0\farcs15$ accuracy of the link between the GSC2 and the International Celestial Reference Frame, we estimate an overall ($1 \sigma$) uncertainty on the \fors\  astrometry of $\delta_{\rm r}\sim0\farcs28$.

\begin{figure*}
{\includegraphics[height=7.5cm,bb=15 190 495 615,angle=0,clip=]{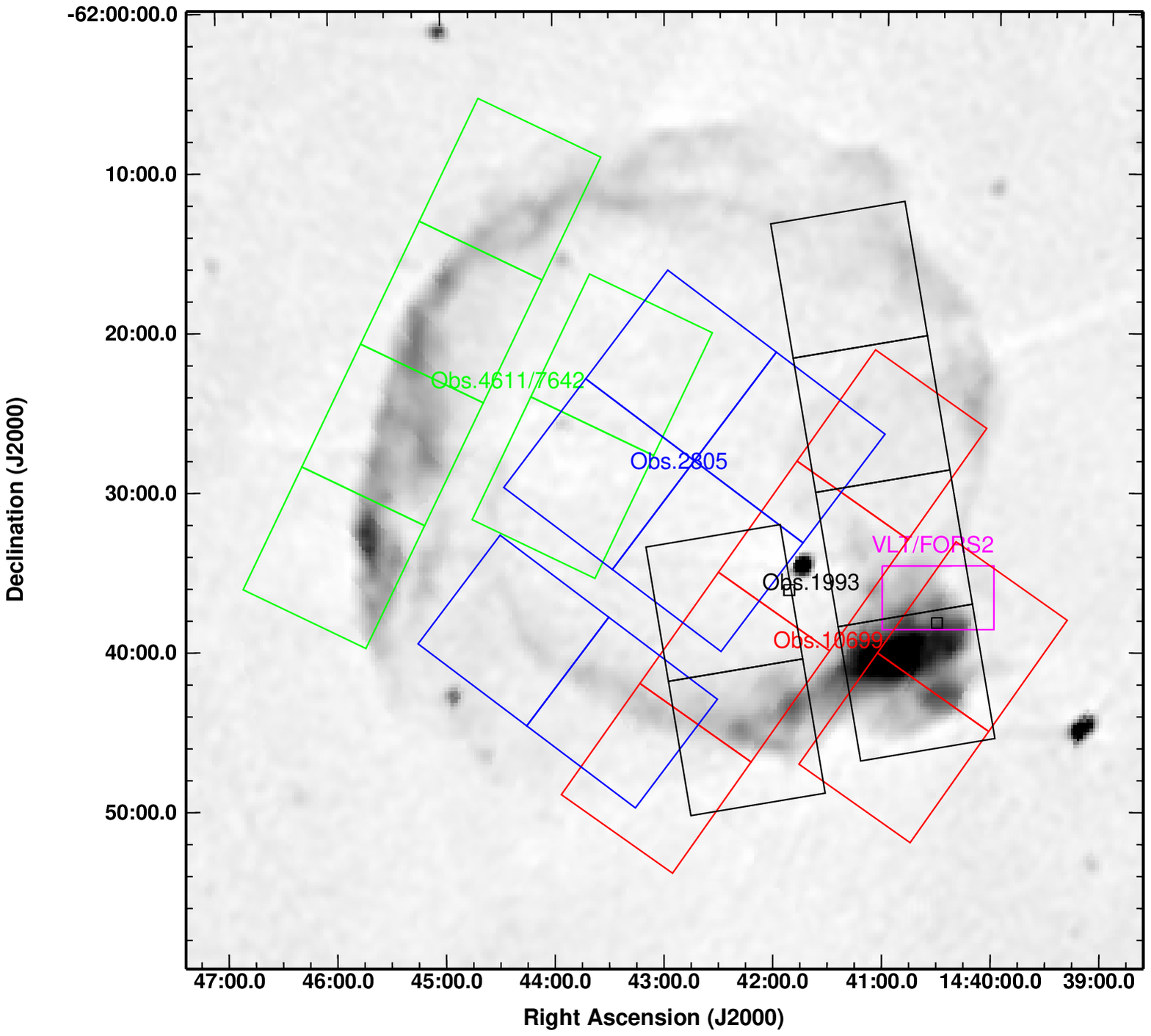}} 
{\includegraphics[height=7.5cm,bb=60 200 540 625,angle=0,clip=]{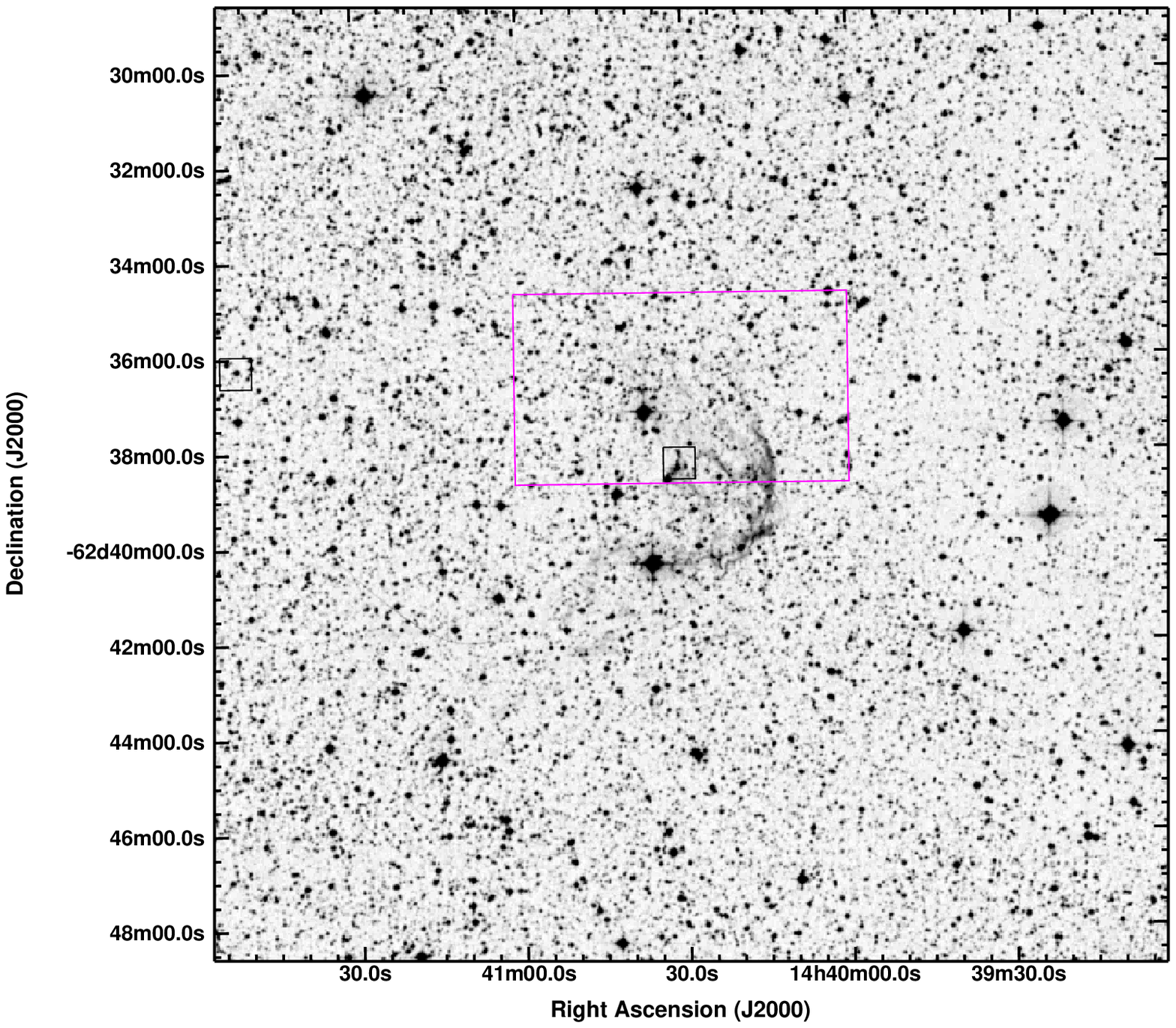}}
{\includegraphics[height=6cm,angle=0]{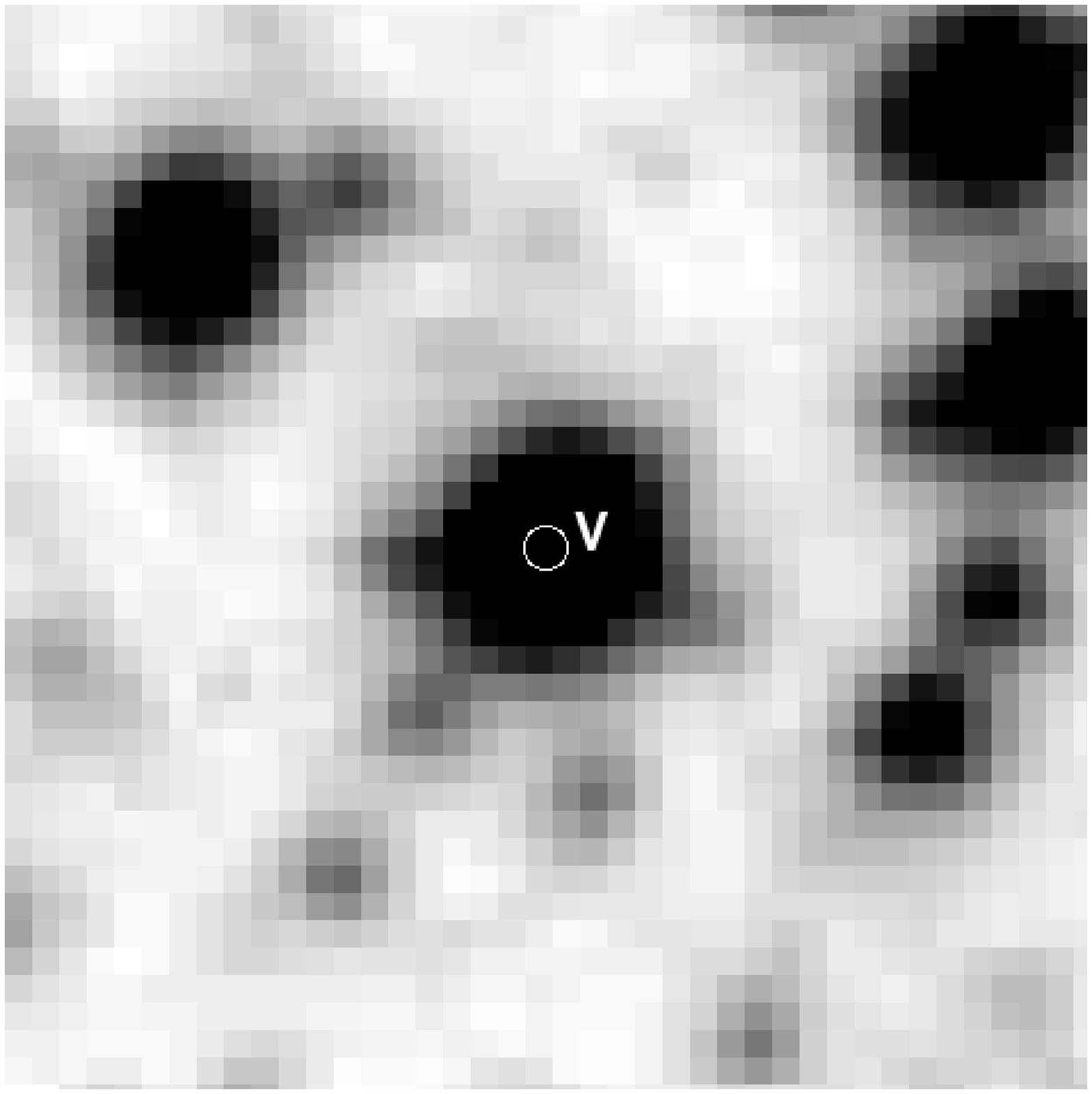}} 
{\includegraphics[height=6cm,bb=38 162 560 684,angle=0,clip=]{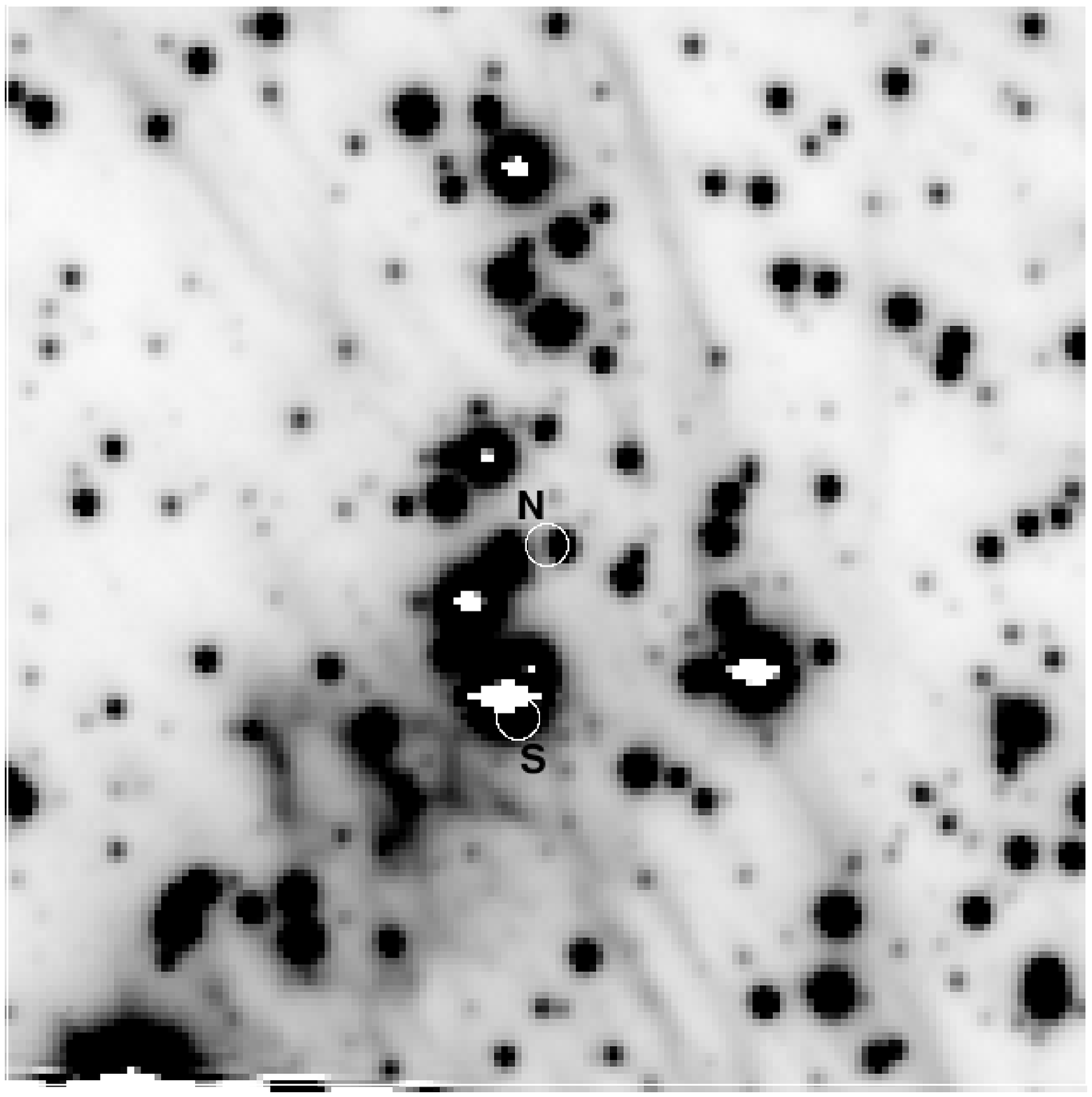}}
\caption{\label{opt} {\em Top left panel:} $60\arcmin \times 60\arcmin$ image of \snr\  from the 843 MHz Sydney University Molonglo Sky Survey (Bock et al.\ 1999). The squares and the rectangle ($7\arcmin\times4\arcmin$) mark the FOV of the \chan/ACIS and VLT observations, respectively. 
{\em Top right panel:} $20\arcmin \times 20\arcmin$ DSS-2 image centred on the RCW\, 86 nebula,  with the FOV of the VLT observation also overlaid.
The two squares ($40\arcsec\times40\arcsec$) mark the DSS-2 ({\em lower left panel}) and \fors\ images ({\em lower right panel}) centred on 
Source V 
and  Source N, 
respectively. The circles ($0\farcs7$) marks their \chan\  positions (Sectn. 3).  The position of Source S  (Gvaramadze \& Vikhlinin 2003) is also shown.}
\end{figure*}

\section{X-ray source optical identification}

Fig.\ 1 (top right) shows a $20\arcmin \times 10\arcmin$ DSS-2 image 
centred on  RCW\, 86, with the \fors\  FOV and the $40\arcsec\times40\arcsec$ regions around  Source V 
and Source N 
overlaid.
We used \chan\ to compute a more precise X-ray position of Source V with respect to that obtained by \ros\ (5\arcsec\ error radius).  We downloaded a short (2 ks)  \emph{Chandra}/ACIS-S observation
(Obs. Id. 10699) from the public \chan\ Science Archive\footnote{\texttt{http://cxc.harvard.edu/cda}}.  We analysed the dataset with the
\chan\ Interactive Analysis of Observations ({\sc CIAO}) software (v. 4.4.2). We extracted an image in the 0.5--6 keV energy range using the original ACIS pixel size ($0\farcs492$). We ran a source detection using the {\tt wavdetect} task with wavelet scales ranging from 1 to 16 pixels, spaced by a factor $\sqrt{2}$ (consistent results were obtained using the {\tt celldetect} task).  We easily detected  Source V and computed its position as  $\alpha =14^{\rm h}  41^{\rm m} 51\fs46$; $\delta   = -62^\circ 36\arcmin 16\farcs5$.  The short integration prevented us to check the accuracy of the \chan\/ACIS-S absolute astrometry by cross-correlating positions of field X-ray sources to optical/infrared astrometric catalogues. Indeed, no  X-ray sources are detected within 4\arcmin\ from the aim point, and poor photon statistics hampers accurate localisation of sources at larger off-axis angle (where PSF blurring becomes significant). We attach to the position of Source V, located very close to the aim point (where the accuracy of the ACIS astrometry is optimal) and detected with a good signal--to--noise (38 photons with virtually no background), the nominal ACIS radial uncertainty of 0\farcs6  (90\% confidence level). 
We repeated the exercise for Source N using a 93 ks \chan/ACIS-S observation (Obs. Id. 1993).
In this case, the much deeper integration allowed us to detect several point X-ray sources in the field, which we used to check the astrometric accuracy. We selected X-ray sources detected at $>4\sigma$ and cross-correlated their positions with the 2 Micron All Sky Survey (2MASS) Catalog (Skrutskie et al.\ 2006).
Unfortunately, we found only one match found within 4\arcmin\ of the aim point. Thus, we included larger off axis regions. Using 9 matches  within 8\arcmin, we assessed that no corrections were needed to the ACIS astrometry (r.m.s. difference of $\sim0\farcs3$  per coordinate). We computed the position of Source N as $\alpha =14^{\rm h}  40^{\rm m} 31\fs11$; $\delta = -62^\circ 38\arcmin 16\farcs7$, with a nominal radial uncertainty of $0\farcs6$  ($90\%$ confidence level). This position is consistent with that measured by Gvaramadze \& Vikhlinin (2003).

Fig.\ 1 (bottom left) shows a $40\arcsec \times 40\arcsec$ zoom of the DSS-2 image centred on the  \chan\ position of Source V.  
The 
source  is positionally coincident  (within the $0\farcs3$ accuracy of the DSS-2 astrometry)  with the bright star 
already proposed by Vink et al.\  (2000) as its potential counterpart. Thus, thanks to the factor of 10 better positional accuracy of \emph{Chandra} with respect to \ros, we can conclude that Source V is associated with this star and cannot be an INS. This means that Source N remains the only possible candidate.
Fig.\ 1 (bottom right) also shows  a $40\arcsec \times 40\arcsec$ zoom of the  \fors\ V-band image around Source N's position. 
The position of 
Source S of Gvaramadze \& Vikhlinin (2003) is also shown. However, 
this was already excluded by these authors as a candidate INS, based on its association with a DSS-2 star (V$\sim 14.3$), saturated in the \fors\ images. 
We found a point source overlapping the revised \emph{Chandra} error circle of Source N, which we propose as its candidate optical counterpart.  We computed its flux  through standard aperture photometry and derived a magnitude $V=20.4 \pm 0.05$.
To quantify the robustness of the association, we computed the probability of a chance coincidence between Source N and its candidate counterpart. We defined it as $P=1-\exp(-\pi\rho r^2)$, where  $r$ is  the matching  radius  (0\farcs7, taking account both the uncertainties of the {\sl Chandra} position and the accuracy of our \vlt\ astrometry), and $\rho$ is  the  density of  stellar objects  in  the \fors\  FOV with magnitudes comparable to the candidate counterpart.  For $\rho  \sim  0.006$ arcsec$^{-2}$, this gives $P\sim 0.008$.
Unfortunately, there are no other images of the field in the ESO archive, 
to obtain colour information on Source N's counterpart.  
Using  a DSS-2 red-band image, Gvaramadze \& Vikhlinin (2003) set an upper limit of $R_F \sim 21$ on Source N, which implies a $(V-R_F) \la -0.6$. 
From the same image
we estimated a $3\sigma$ upper limit of $R_F \sim 19$ and  $(V-R_{F}) \la 1.4$.  However, due to the high sky background ascribed to a group of bright stars at $\la 3\arcsec$ from  Source N (Fig.\ 1, bottom right), unresolved in the DSS-2 image, these upper limits are affected by a large uncertainty.  This is also true for those obtained from 2MASS images.
Thus, we cannot derive tight constraints on the colours of Source N's counterpart.

\section{X-ray spectral and timing analysis}

We used the available X-ray and optical data to determine whether  Source N can be 
an INS.
Deep observations of  Source N's field have been obtained both by \chan\ (Obs. Id. 1993; 93 ks) and \xmm\ (Obs. Id. 0504810401; 72 ks). Unfortunately, due to the positional coincidence with a bright X-ray filament of the SNR,
which dominates the emission below 2 keV,  Source N is possibly detected only at energies above $\sim$ 2 keV in the \xmm\ observation, which makes such data not suitable for the spectral analysis.
Thus, we used the  \chan\/ACIS-S data to characterise the X-ray spectrum of Source N. We extracted counts for both the point source and background using the {\tt specextract} script in {\sc CIAO}.
We selected source photons from circular apertures with a 2\arcsec\  radius and background photons  from two circles with a radius of 5\arcsec.
As done by Kaplan et al.\ (2004), we compared the source counts in the low (0.3--2 keV) and hard (2--8 keV) energy bands and computed the hardness ratio $HR=(C_{H}-C_{L})/(C_{H}+C_{L})=-0.21\pm0.08$, where $C_{H}= 161\pm13$ and $C_{L}=245\pm25$ are the counts in the hard and low energy bands, respectively. The  $HR$  is compatible, within the observed scatter, with those of the X-ray sources detected by Kaplan et al.\ (2004) at the centre of \snr. For Source V we obtained $HR=-0.95\pm0.34$, which is also compatible with the range of values for these sources. In particular the $HR$ is consistent with Source V being a star, as implied by its optical identification.
We did a more qualitative spectral modelling of Source N using  {\sc XSPEC} (v. 12.7).  The spectrum of Source N is well described (reduced $\chi^2=1.00$, 28 d.o.f.) by a power law (PL) with photon index $\Gamma=2.0\pm0.2$, 
absorbed by an hydrogen column density N$_{\rm H}=(7\pm2)\times10^{21}$ cm$^{-2}$.  We note that our PL spectral index is consistent with that measured by  Gvaramadze \& Vikhlinin (2003), although they fixed the value of $N_{\rm H}=1.5 \times 10^{21}$ cm$^{-2}$. An equally good fit  (reduced $\chi^2=1.00$, 28 d.o.f.) can be obtained using a simple blackbody (BB) spectrum with kT=$0.83\pm0.05$ keV, absorbed by a low column density N$_{\rm H}<10^{21}$ cm$^{-2}$.   An  optically thin Raymond-Smith (RS) thermal plasma model ({\em raymond} in {\sc XSPEC}) with kT$=4.8_{-1.5}^{+2.5}$ keV, abundance $<1.2$ solar values and an absorbing column N$_{\rm H}=(5\pm2)\times10^{21}$ cm$^{-2}$ also fits well the data (reduced $\chi^2=1.01$, 27 d.o.f.).   We also fitted the \chan\ spectrum with two-component models, as  done by  Gvaramadze \& Vikhlinin (2003), but the fit quality does not improve noticeably with respect to single-component ones. Thus, from the X-ray spectrum alone we cannot unambiguously determine the nature of Source N.
To further investigate the possibility that Source N is an INS, we searched for periodicity in the \xmm/EPIC-pn  data. However, since these were taken in {\em full frame} mode (frame time of 73.4 ms), the periodicity search is only sensitive to periods $>$146.8 ms.  We used tools in the {\sc SAS} package  (v. 11.0) to extract X-ray events from a 20 \arcsec\ circle centred at the computed \chan\  position of Source N and converted their times of arrival (TOAs) to the Solar System barycentre. We performed a search for coherent pulsations in different energy bands (mainly above 2 keV)
using the {\tt powspec} tool in  
{\sc FTOOLS}, but we found no significant signals. Unfortunately, both the limited number of source counts against the high background produced by the bright X-ray filament of the SNR and the contamination of Source S, only 6 \arcsec\ away, did not allow us to obtain constraining limits on the pulsed fraction.  

\section{Discussion and conclusions}

To establish the nature of Source N, we evaluated whether the  flux of its optical counterpart is compatible with what expected for an INS. For the estimated age of \snr\ ($\sim 1800$ yrs) and an optical luminosity comparable to the Crab pulsar, we would  expect a flux of
$V\sim 16.6$--21.1, after re-normalising for the \snr\ distance (1--2.8 kpc) and interstellar extinction $A_V\sim2.8$--5, inferred upon the N$_{\rm H}$
derived from the PL fit to the \chan\ spectrum using the relation of Predhel \& Schmitt (1995).
These values are compatible with the flux of Source N's counterpart ($V=20.4$), at least for the highest values of the interstellar extinction and distance. Alternatively, Source N might be a CCO. For a rotational energy loss rate $\dot{E}<10^{36}$ erg s$^{-1}$ (
see De Luca et al.\ 2012 and references therein),  the expected optical luminosity would then be at least two orders of magnitude lower than the Crab pulsar, yielding a flux $V\ga23.5$.
We also verified whether the X-ray--to--optical flux ratio of Source N is compatible with that of an INSs.
According to the PL model, the observed flux in the 0.3--8 keV energy range is $4.2_{-0.5}^{+0.2}\times10^{-14}$ erg cm$^{-2}$ s$^{-1}$, corresponding to an unabsorbed flux $F_{\rm X} = 8.5_{-0.9}^{+0.4}\times10^{-14}$ erg cm$^{-2}$ s$^{-1}$.  The unabsorbed optical flux of Source N's  optical counterpart, computed upon the corresponding N$_{\rm H}$,  is $F_{\rm opt} = (0.28-2.1) \times 10^{-12}$ erg cm$^{-2}$ s$^{-1}$. Thus, the unabsorbed X-ray--to--optical flux ratio of Source N  would be $F_{\rm X}/F_{\rm opt} \sim 0.036$--0.32, by far lower than expected for an INS,  for which this ratio is usually of 1000, or higher (e.g., Mignani 2011). Assuming other best-fitting X-ray spectral models also gives very modest values of the $F_{\rm X}/F_{\rm opt}$ ratio. 
This implies that Source N must be a different type of X-ray source, maybe an AGN,  which would be compatible with its $F_{\rm X}/F_{\rm opt}$ ratio (e.g. la Palombara et al.\ 2006) and its possible PL spectrum. Spectroscopy of its optical counterpart is needed to determine the nature of Source N.
Thus, from their associations with bright optical counterparts, we conclude that neither Source V 
nor Source N 
are INSs. No other unidentified point-like X-ray sources which can be considered possible INS candidates have been discovered by \chan\ surveys of \snr.  The available \chan\ observations (Fig.\ 1, top left) cover a large fraction of the SNR. In particular, the most likely locations where the compact remnant of the SN explosion is expected to be found, i.e. the central regions of the SNR and the RCW\, 86 nebula, have been deeply scrutinised (e.g., Kaplan et al.\ 2004; Gvaramadze \& Vikhlinin 2003). Thus, it is unlikely that potential INS candidates were missed.   We conclude that the lack of an identified INS  supports the conclusion  that \snr\ was not born after a Type II SN explosion (Williams et al.\  2011). 

\section*{Acknowledgments}RPM thanks Dave Green for useful comments on \snr.

\label{lastpage}

\end{document}